\documentclass{ws-ijmpd}

\begin{document}

\markboth{B.J. Ahmedov and V.G. Kagramanova}
{Electromagnetic Effects in Superconductors in Gravitational Field}

%%%%%%%%%%%%%%%%%%%%% Publisher's Area please ignore %%%%%%%%%%%%%%%
%
\catchline{}{}{}{}{}
%
%%%%%%%%%%%%%%%%%%%%%%%%%%%%%%%%%%%%%%%%%%%%%%%%%%%%%%%%%%%%%%%%%%%%

\title{Electromagnetic Effects in Superconductors\\ in Stationary
Gravitational Field}
\author{\footnotesize B.J. AHMEDOV}

\address{Institute of Nuclear Physics and
    Ulugh Beg Astronomical Institute\\ Astronomicheskaya 33,
    Tashkent 700052, Uzbekistan\\
   IUCAA, Post Bag 4 Ganeshkhind,
411007 Pune, India\\
    ahmedov@astrin.uzsci.net}

\author{V.G. KAGRAMANOVA}

\address{Uzbekistan National University,
    Tashkent 700095, Uzbekistan         \\
         IUCAA,
         Post Bag 4 Ganeshkhind,
411007 Pune, India\\
kavageo5@rambler.ru }

\maketitle

\begin{history}
\received{9 September 2004}
%\revised{Day Month Year}
%\accepted{Day Month Year}
%\comby{(xxxxxxxxxx)}
\end{history}

%\pub{Received (12 12 2003)}{Revised (00 00 2003)}
%\pub{Received (0 0 0)}{Revised (0 0 0)}

\begin{abstract}
The general relativistic modifications to the resistive state in
superconductors of second type in the presence of a stationary
gravitational field are studied. Some superconducting devices that
can measure the gravitational field by its red-shift effect on the
frequency of radiation are suggested. It has been shown that by
varying the orientation of a superconductor with respect to the
earth gravitational field, a corresponding varying contribution to
AC Josephson frequency would be added by gravity. A magnetic flux
(being proportional to angular velocity of rotation $\Omega$)
through a rotating hollow superconducting cylinder with the radial
gradient of temperature $\nabla_r T$ is theoretically predicted.
The magnetic flux is assumed to be produced by the azimuthal
current arising from Coriolis force effect on radial
thermoelectric current. Finally the magnetic flux through the
superconducting ring with radial heat flow located at the
equatorial plane interior the rotating neutron star is calculated.
In particular it has been shown that nonvanishing magnetic flux
will be generated due to the general relativistic effect of
dragging of inertial frames on the thermoelectric current.

\end{abstract}

%\ccode{PACS Nos.: 04.20.-q; 04.40.-b; 04.80.Cc}


\section{Introduction}

Although the effects of gravitational field on the electromagnetic
properties of superconductors have been widely discussed by a
number of authors (see, for example, \cite{dw}-\cite{a99}) there
exist unconsidered problems which are of physical and
astrophysical interest. To the best of our knowledge theoretical
investigation of the effects of general relativity on
electrodynamics of laboratory II-type superconductors has not been
done. This subject is very interesting, on one hand, from point of
view to detect weak gravitational effects in the solar system
since superconductors provide sensitive and accurate measurements.

On other hand it is important for astrophysics of
magnetized relativistic compact objects since the core of neutron
stars forms a
matter in the superconducting state of II-type~\cite{bpp69}
arising from the
realistic estimation that the superconducting protons coherence
length ($2-6 fm$) for stellar matter is typically much smaller
than the London screening length ($100-300 fm$). Magnetic flux
vortices are the result of the type-II superconductivity
in the inner crust and core region. The magnetic field of the neutron
star is confined into individual vortices of flux $\Phi_0=\pi\hbar
c/e=2\times 10^{-7}G\cdot cm^2$. The number density of such
vortices is $n_v=B/\Phi_0\approx 10^{19} B_{12}cm^{-2}$, where
$B_{12}=B/10^{12} G$. The knowledge of the electromagnetic
properties of flux vortices in gravitational field may be relevant
for understanding some observable phenomena of pulsars~\cite{sls00}.
However, we should mention here that in recent papers (see, for example,
\cite{bmz04}) it was discussed inconsistency of the standard picture of
the neutron star core, composed of a mixture of a neutron superfluid and
a proton type-II superconductor
with observations of long period precession in isolated pulsars.

The present work is the sequel of our previous
investigation~\cite{a99,a98} on gravitational and rotational effects
on electromagnetic properties of conductors and superconductors.
In our previous paper~\cite{a99} we did not consider the electromagnetic
properties of type-II superconductors in the gravitational field.
One of the purpose of the paper is to study gravitational effects on
type-II superconductivity. The second purpose of the paper is to
examine a question on generation of magnetic field in superconductors
arising from the effects of gravity or inertia. Since the mechanism
of the generation of neutron star magnetic fields is under big
astrophysical interest, we apply the discussed mechanism to
the interior of slowly rotating neutron star in the simple toy model.

The paper is organized as follows. In the section~\ref{sup2} we
consider the general relativistic modifications to the electromagnetic
properties of II-type superconductor when
the Lorentz force created by transport current flowing in the
superconductor exceeds the pinning force and the vortices started to move
laterally to the current (This stage is called as resistive one
due the appearance of the resistance from the energy
dissipation.). This generalizes the resisitive state in II-type
superconductors when the gravitational field is present. In addition,
in subsection~\ref{redshift}
 the weak general relativistic contribution on the frequency of
radiation from II-type superconductor with transport current in an
applied magnetic field is calculated.
More concretely we propose here an experiment where the AC
Josephson effect in a superconductor of II-type has a combined
electric and gravitational origin.

The magnetic flux through superconducting hollow cylinder arising
from interplay between thermally generated electric current and
uniform rotation is obtained in section~\ref{rotation}. In
section~\ref{star} new possible mechanism for magnetic field
production inside a rotating relativistic star is discussed. The
last section~\ref{conclusion} summarizes our results.

We use here a space-like signature $(-,+,+,+)$ and a system of
units in which $c=1$ (unless explicitly shown otherwise for
convenience). Greek indices are taken to run from $0$ to $3$.

\section{Second type superconductor in resistive state
in presence of gravitational field}
\label{sup2}

Suppose that superconducting sample is placed in the gravitational
field being assumed to be stationary with respect to the
superconductor as whole, that is there exist a timelike Killing
vector $\xi_{(t)\alpha}$ being parallel to the four velocity of
the superconductor $u^\alpha$. On defining
$\Lambda=-\xi_{(t)}^\alpha\xi_{(t)\alpha}$, Killing equation
$\xi_{(t)\mu ;\nu} + \xi_{(t)\nu ;\mu}=0$ implies that
$\partial_\alpha \Lambda^{1/2} = \Lambda^{1/2}w_\alpha$, where
$w_\alpha=u_{\alpha;\beta}u^\beta$ is the absolute acceleration of
superconductor. The electric and magnetic fields as seen by
observers who are at rest with respect to the superconductor are
$E_\alpha=F_{\alpha\beta}u^\beta$ and $B_\alpha=(1/2)
\eta_{\alpha\beta\mu\nu} F^{\beta\mu}u^\nu$, where
$F_{\alpha\beta}=A_{\beta,\alpha}-A_{\alpha,\beta}$ is the field
tensor, $A_\alpha$ is the four-potential of electromagnetic field,
$\eta_{\alpha\beta\mu\nu}=\sqrt{-g}\epsilon_{\alpha\beta\mu\nu}$
is the pseudo-tensorial expression for the Levi-Civita symbol
 $\epsilon_{\alpha\beta\mu\nu}$, $g\equiv det|g_{\alpha\beta}|$.

Electromagnetic field applied to the sample is supposed to be
stationary and $\pounds_{\xi_t} F_{\alpha\beta}=0$ (which is
equivalent to $\pounds_{\xi_t} A_\alpha$ in the Lorentz gauge),
where $\pounds_{\xi_t}$ denotes the Lie derivative with respect to
vector field $\xi_{(t)}^\alpha$.

For laboratory II-type superconductor the London penetration depth
$\lambda$ is bigger than the coherence length $\xi$
($\lambda\gg\xi$) and external magnetic field ${\bf B}_e$
penetrates the superconductor as an array of vortex lines but in
the space between vortices the material remains superconducting.
Assume that the massive superconductor is in resistive state in an
applied magnetic field. For the simplicity we will consider
vortices which form quadratic lattice (see Fig. 1). Their axes are
aligned along the magnetic field lines.

\begin{figure}
\hspace*{0.5cm}
\centerline{ \psfig{figure=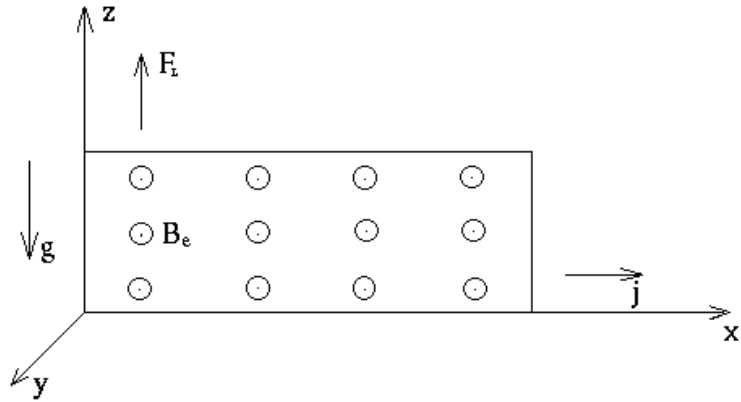,width=4.0in}}
\vspace*{8pt}
\caption{ II-type superconductor with transport
current ${\mathbf j}$ embedded in the external magnetic field
${\mathbf B}_e$ and gravitational field ${\bf g}$. A Lorentz force
${\mathbf F}_L$ acts on the flux tubes and moves them in the $z$
direction.}
\end{figure}

If the superconductor is in the mixed state and transport current,
created by the external source, flows in the direction being
perpendicular to the vortices then the Lorentz force acting on the
vortices appears. The vortex state has zero electrical resistivity
if the flux tubes are prevented from moving in response to an
external magnetic field. However due to the pinning of vortices it
is necessary to create the finite electric current (called as
critical one) for motion of vortices.

Transport current creates Lorentz force under which the whole
vortice structure moves with relative velocity $v_{(v)}^\alpha$
defined through decomposition formula:
\begin{equation}
u^{\alpha}_{(v)}=\frac{u^\alpha+v_{(v)}^\alpha}{\sqrt{1-v_{(v)}^2}}
\approx u^\alpha+v_{(v)}^\alpha \ .
\end{equation}
Hereafter we consider slow motion case and neglect terms being quadratic
in the relative velocity $v_{(v)}^\alpha$.

The flux tubes must migrate perpendicular to the direction of
current and perpendicular to the magnetic field under the
influence of Lorentz force, i.e. the normal core regions move also
as the electrons drifting in an external electric field. Such a
drift or current causes ohmic heat loss. Thus migration of the
flux tubes through the superconductor causes the occurrence of
dissipation and the sample exhibits an electrical resistance.

Now we calculate the dissipation of electromagnetic energy due to
the motion of magnetic field inside the superconductor. The
electromagnetic energy-momentum tensor is
\begin{equation}
\label{energymomentum}
T^{\alpha\beta}_{(em)}=\frac{1}{4\pi}\left(F^{\alpha\sigma}
F^{\beta}_{.\sigma}-\frac{1}{4}g^{\alpha\beta}
F^{\mu\nu}F_{\mu\nu}\right)\ .
\end{equation}

According to the first pair of Maxwell equations
\begin{equation}
F_{\mu\nu;\alpha}=-F_{\nu\alpha;\mu}-F_{\alpha\mu;\nu}
\label{max1}
\end{equation}
the divergency of energy-momentum tensor (\ref{energymomentum})
\begin{equation}
\label{divergency}
T^{\beta}_{(em)\alpha;\beta}=\frac{1}{4\pi}\left(
F_{\alpha\sigma;\beta}F^{\beta\sigma}
+F_{\alpha\sigma}F^{\beta\sigma}_{;\beta}-
\frac{1}{2}F^{\mu\nu}F_{\mu\nu;\alpha}\right)\
\end{equation}
takes form
\begin{equation}
T^{\beta}_{(em)\alpha;\beta}=\frac{1}{4\pi}
F_{\sigma\alpha}F^{\sigma\beta}_{;\beta} \ .
\end{equation}

Now use of the second pair of Maxwell equations
\begin{equation}
F^{\alpha\beta}_{;\beta}=4\pi J^\alpha
\label{max2}
\end{equation}
gives
\begin{equation}
\label{divergency1}
T^{\alpha\beta}_{(em);\beta}=F^{\sigma\alpha}J_{\sigma} \ ,
\end{equation}
where the four current $J_\alpha=\rho_e u_\alpha+j_\alpha$,
$\rho_e$ is the electric charge density, $j_\alpha$ is the
conduction current.

Now if we scalarly multiply the equation (\ref{divergency1}) to
4-velocity of the superconductor $u_\alpha$ we get the formula
\begin{equation}
\label{joule} u_\alpha T^{\alpha\beta}_{(em);\beta}= u_\alpha
F^{\sigma\alpha}J_{\sigma}= E^\sigma j_\sigma
\end{equation}
for the electromagnetic energy change along the world lines of
superconducting medium.

The conversion of electromagnetic energy to heat as a result of
the migration of flux vortices will basically take place as the
occurrence of local electric field $E_\alpha$. The migration of
flux vortices causes the magnetic field to change with time which
results in an electric field
\begin{equation}
\label{Ealpha}
E_{}^\alpha=\eta^{\alpha\beta\mu\nu}u_\beta
v_{(v)\mu}B_{\nu}\ .
\end{equation}

The electric field~(\ref{Ealpha}) then accelerates the unpaired electrons
which can pass their energy taken from the electric field, on to
the lattice and, hence, produce a heat. The proper resistance of
superconductor $\rho_f$, which appears due to the lateral motion
of magnetic flux with respect to the transport current, is called
as flux-flow resistance and differs from the ohmic resistance
$\sigma^{-1}$ for normal electrons.

\subsection{The gravitational red-shift correction to the frequency of
radiation from II-type superconductor  in resistive state}
\label{redshift}

Now we will show a similarity between Josephson junction in
general relativistic AC generation state~\cite{an1}-\cite{a99} and
superconductor of II-type in the resistive state in the presence
of stationary gravitational field.

Left hand side of equation (\ref{Ealpha}) can be written as
\begin{equation}
E_\alpha= -\Lambda^{-1/2}A_{\alpha,\beta}\xi^\beta_{(t)}+
          \Lambda^{-1/2}A_{\beta,\alpha}\xi^\beta_{(t)}\ .
\end{equation}

Taking into account that 4-potential $A_\alpha$ must satisfy the
equation $\pounds_{\xi_t} A_{\alpha}=0$, one can get the following
expression
\begin{equation}
E_\alpha=\Lambda^{-1/2}\varphi_{,\alpha}\ , \qquad
\varphi\equiv A_\rho\xi^\rho_{(t)}\ . \label{Ealpha1}
\end{equation}

Substituting (\ref{Ealpha1}) into (\ref{Ealpha})~\footnote{We
neglect here the effect of dynamical part of gravitational field
on motion of flux vortices since we make the concrete analysis in
the Schwarzschild space-time.}
\begin{equation}
\Lambda^{-1/2}\varphi_{,\alpha}= \eta_{\alpha\beta\mu\nu}u^\beta
v_{(v)}^{\mu}B^{\nu} \ . \label{lambda}
\end{equation}

Suppose that the type-II superconductor is in the Schwarzschild
space-time
\begin{equation}
ds^2=-\left(1-\frac{2M}{r}\right)dt^2+
\left(1-\frac{2M}{r}\right)^{-1}dr^2 +r^2d\theta^2+r^2\sin^2\theta
d\phi^2 \ ,
\end{equation}
where the timelike Killing vector can be chosen so that $\Lambda
=1-2M/r$. If the curvature effects are negligible, then the
apparatus may be regarded as having an acceleration, $g$, relative
to a local inertial frame, and thus, $\Lambda =
\left(1+2gz/c^2\right)^2$, where $z$ is the height above some
fixed point.

Consider a rectangular superconducting plate carrying an electric
current being parallel to its plane. It is in vertical
gravitational field $g$ and held in an applied perpendicular magnetic
field $B$. If $\overline{V}$ is a potential
difference on the length of period $a$ in $z$ direction then
equation~(\ref{lambda}) takes form
\begin{equation}
\Lambda^{-1/2}\overline{V}=a\left(Bv_{v}\right) \ .
\label{V}
\end{equation}

It is clear that the displacement of vortices is translationally
symmetric with respect to the period $a$. Thus one could expect
there is an alternating component  of voltage $\overline{V}$ with the
frequency
\begin{equation}
\omega=2\pi\frac{v_{(v)}}{a}\ .
\label{omega}
\end{equation}

Inserting (\ref{omega}) into (\ref{V}) gives
\begin{equation}
\Lambda^{-1/2}\overline{V}=\frac{\omega(Ba^2)}{2\pi c}\ .
\label{V1}
\end{equation}

Taking into account that $Ba^2$ is the magnetic flux connected
with the single vortice, i.e. flux quantum is $\Phi_0=\pi\hbar
c/e$ we can rewrite (\ref{V1}) in the form
\begin{equation}
\Lambda^{-1/2}\overline{V}=\frac{\hbar\omega}{2e}\ .
\label{V2}
\end{equation}

As a consequence of (\ref{lambda}) and (\ref{V2}), an alternating
current of frequency
\begin{equation}
\omega=
    \frac{2e}{\hbar}\Lambda^{-1/2}{\overline{V}}
\label{om}
\end{equation}
is produced.
Consequently $\omega =\omega_0\left(1-gz/c^2\right)$, where
$\omega_0$ is a constant.

According to  (\ref{om}) the
frequency of the alternating
current depends on the altitude in the gravitational field and
hence frequences $\omega_{1}$ and $\omega_{2}$ for the heights above
some points $z_{1}$ and $z_{2}$ ($z=z_{2}-z_{1}; z_{1}<z_{2}$) are connected
through
\begin{equation}
\omega(z_{2})=\omega(z_{1})(1-gz/c^2) \ .
\end{equation}
Using typical numbers for laboratory experiments in the
Earth's gravitational field
as voltage $\overline{V}=100V$, height $z=10cm$ and gravitational
acceleration $g=9.8\times 10^2 cm\cdot s^{-2}$
one can get the
general relativistic red-shift for the frequency
\begin{equation}
\omega=
\frac{2e}{\hbar}\frac{gz}{c^2}{\overline{V}} \approx{3.3Hz}\ .
\end{equation}
This is the gravitational redshift of
vortices, and superconductors of II-type in principle could
provide a test of gravitational field with help of the Josephson
effect.

\section{Effect of uniform rotation on a hollow superconducting
cylinder with radial heat flow}
\label{rotation}

It was obtained in~\cite{a99} that an azimuthal temperature
gradient applied to a hollow bimetallic superconducting cylinder
(ring) furnishes a small magnetic field consisting of a
contribution due to the trapped flux $n\Phi_0$ ($n$ is the number
of initially trapped quanta)
%, $\Phi_0=\pi\hbar c/e=2\times
%10^{-7}G\cdot cm^2$ is the magnetic flux quantum)
plus the field induced by the thermal current
$j_{(s)\nu}=\Lambda^{-1/2}\sigma\beta \partial_\nu{\tilde T}$
($\beta$ is the thermoelectric coefficient, $\sigma$ is the
conductivity of the normal component).

Here taking into account thermoelectric effects in
superconducting state~\cite{a99} we extend to the superconducting
case our result~\cite{a98} on the rotational effects on the
radial conduction current of thermoelectric origin. Consider now
bimetallic hollow cylinder of outer radius $r_2$ and internal
radius $r_1$ in the presence of radial heat flow (see Fig. 2).

\begin{figure}
\hspace*{0.5cm} \centerline{
\psfig{figure=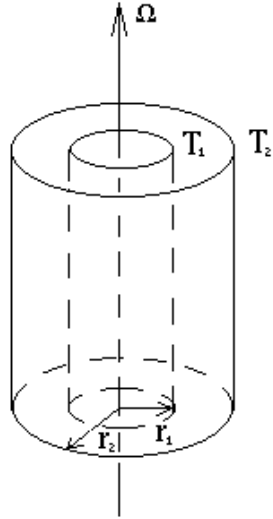,width=6.0in}} \vspace*{3pt} \caption{
Superconducting cylinder of outer radius $r_2$ and internal radius
$r_1$ rotating with angular velocity $\Omega$ along its axis.
$\Delta T=T_2-T_1$, where $T_2(T_1)$ is the hot (cold) layer
temperature.}
\end{figure}

As was shown in~\cite{a99}, in the bulk of a superconductor the
total thermoelectric current
$j^\alpha=j_{(s)}^\alpha+j_{(n)}^\alpha$ vanishes, i.e. the normal
current density $j_{(n)}^\alpha$ is cancelled locally by a
counterflow of supercurrent density $j_{(s)}^\alpha$, so that
\begin{eqnarray}
\label{cancel}
j_{(n)\alpha} =
    \Lambda^{-1/2}\sigma\beta\partial_\alpha{\tilde T}- \sigma
    R_H\left(F_{\nu\alpha}+u_\alpha u^\sigma F_{\nu\sigma}\right)
    j^{\nu}_{(n)}+ \sigma R_{gg}j^{\beta}_{(n)} A_{\alpha\beta}=
\nonumber\\ \nonumber\\
-j_{(s)\alpha} =
    \frac{2n_se}{m_s}\left[\hbar\partial_\alpha\vartheta
    -\frac{2e}{c}A_\alpha\right]\ ,
\end{eqnarray}
where $A_{\alpha\beta}=u_{[\alpha,\beta]}+u_{[\beta}w_{\alpha ]}$
is the relativistic rate of rotation, $n_s$ and $m_s$ represent
the density and mass of Cooper pairs, $\vartheta$ is the phase of
superconducting wavefunction. $R_H$ is the Hall constant and
$R_{gg}$ is the galvanogravitomagnetic coefficient (
$R_{gg}\approx0.8\times 10^{-22} s^2$, see for
details~\cite{a99a}). Thus the initial electric current
(initiating a real radial thermoelectric current) could be assumed
to be $j^{\hat r}_{(n)}=\sigma\beta \partial_r{T}$ (the
orthonormal components are hatted).

In the rotating frame of reference (with angular velocity $\Omega$)
\begin{equation}
ds^2=-\left(c^2-\Omega^2r^2\right)dt^2+2\Omega r^2d\phi dt+
dr^2+r^2d\phi^2+dz^2 \
\end{equation}
the radial electric current will be affected by the Coriolis force
and therefore an azimuthal current
\begin{equation}
\label{azimuth} j_{(n)}^{\hat\phi} = -j_{(s)}^{\hat\phi} =
R_{gg}\sigma^2\beta\left(\nabla_rT\right)\Omega
\end{equation}
in the bulk of the superconductor is generated.

Then integrating the equation~(\ref{cancel}) along the contour,
which is in the bulk of the superconductor one can easiely get
\begin{equation}
\Phi_b=n\Phi_0+\frac{m_sc}{4n_se^2}
R_{gg}\sigma^2\beta\left(\nabla_rT\right)2\pi\Omega \ .
\end{equation}
Thus the magnetic flux due to the Coriolis force acting on
superconducting current is expected to be $\Phi_c=\left({\pi
m_sc}/{2n_se^2}\right)
R_{gg}\sigma^2\beta\left(\nabla_rT\right)\Omega$.
Using the typical numbers for conductivity of normal element
$\sigma=9\times10^{16} s^{-1}$, thermoelelectric coefficient
 $\beta=10^{-5} V\cdot K^{-1}$,
angular velocity $\Omega=2\pi\times10^{4} s^{-1}$,
concentration of superconducting electrons
$ n_{s}=10^{23} cm^{-3}$
and $\nabla_rT=10 K\cdot cm^{-1}$ one can evaluate
the additional magnetic flux as
$\Phi_c=5\times10^{-13} G\cdot cm^2$.

\section{Magnetic flux through superconducting ring inside a
rotating neutron star}
\label{star}

Finally, as we noted the mechanism of magnetic field production
discussed above may be relevant to the problem of origin of magnetic
fields in rotating neutron stars whose substance is expected to be
superconducting at high densities.
Assume that the fields interior star are stationary and
axisymmetric, and ignore the self-gravity of the electromagnetic
field. This is realistic assumption since even for the magnetized
star with high magnetic field the electromagnetic field energy is
negligible with compare to the rest mass energy (see, for example,~
\cite{ra04}).

Since the angular momentum $a$ of the neutron star will not be zero
in general, the interior background geometry is given by the metric
at the first order in the angular
momentum in a coordinate system
$(ct,r,\theta,\phi)$ (see, for example,~\cite{ht68}) as follows
\begin{equation}
\label{slow_rot}
ds^2 = -e^{2 \Phi(r)} dt^2 + e^{2
    \Lambda(r)}dr^2 - 2 \omega (r) r^2\sin^2\theta dt d\phi +
    r^2 d\theta ^2+ r^2\sin^2\theta d\phi ^2 \ ,
\end{equation}
where $\omega(r)$ can be interpreted as the angular
velocity of a free falling (inertial) frame and is also
known as the Lense-Thirring angular velocity. The radial
dependence of $\omega$ in the region of spacetime
internal to the star has to be found as the solution of
the differential equation
\begin{equation}
\label{omg_r_int} \frac{1}{r^3} \frac{d}{dr}
    \left(r^4 {\bar j} \frac{d {\bar \omega}}{dr}\right)
    + 4 \frac{d {\bar j}}{dr} {\bar \omega} = 0 \ ,
\end{equation}
where we have defined
\begin{equation}
\label{jbar} {\bar j} \equiv e^{-(\Phi + \Lambda)} \ ,
\end{equation}
and where
\begin{equation}
\label{omegabar} {\bar\omega} \equiv \Omega -\omega \ ,
\end{equation}
is the angular velocity of the fluid as measured from the local
free falling (inertial) frame.

In the vacuum region of spacetime
external to the star, on the other hand, $\omega(r)$ is given by
the simple algebraic expression
\begin{equation}
\label{omg_r_ext} \omega (r)\equiv
\frac{d\phi}{dt}=-\frac{g_{0\phi}}{g_{\phi\phi}}=
    \frac{2J}{r^3} \ ,
\end{equation}
where $J=I(M,R)\Omega$ is the total angular momentum of metric
source as measured from infinity and $I(M,R)$ its momentum of
inertia. Outside the star, the metric (\ref{slow_rot}) is
completely known and explicit expressions for the other metric
functions are given by
\begin{equation}
e^{2 \Phi(r)} \equiv \left(1-\frac{2 M}{r}\right)
    = e^{-2 \Lambda(r)} \ , \hskip 1.0cm
    r > R \ ,
\end{equation}
where $M$ and $R$ are the mass and radius of the star as measured
from infinity.

The four-velocity components of stellar medium are given by
\begin{equation}
\label{u_s/c} u^{\alpha}\equiv
    e^{-\Phi(r)}\bigg(1,0,0,0\bigg) \ ;
    \hskip 2.0cm
u_{\alpha}\equiv
    e^{\Phi(r)}\bigg(- 1,0,0,-\omega r^2\sin^2\theta
    e^{-2\Phi(r)} \bigg) \ .
\end{equation}

    In the coordinate system (\ref{slow_rot}) and
with the electromagnetic fields referred to the medium
(\ref{u_s/c}), the Ohm's law (\ref{cancel}) for the normal
component of electric current takes form
\begin{eqnarray}
\label{ohm_1}
j_{(n)r}&=&\sigma E_r+\sigma R_{gg}j^\phi_{(n)}r\sin^2\theta e^{-\Phi}
\left\{-\omega +\omega r\Phi_{,r}-\frac{\omega 'r}{2}\right\}
+\sigma\beta e^{-\Phi}\left(e^\Phi T\right)_{,r}\ ,
\\ \nonumber\\
\label{ohm_2}
j_{(n)\theta}&=&\sigma E_\theta+\sigma R_{gg}j^\phi_{(n)}\omega r^2
\sin\theta\cos\theta e^{-\Phi}+\sigma\beta T_{,\theta}\ ,
\\ \nonumber\\
\label{ohm_3}
j_{(n)\phi}&=&\sigma E_\phi
+\sigma R_{gg}j^r_{(n)}r\sin^2\theta e^{-\Phi}
\left\{\omega -\omega r\Phi_{,r}+\frac{\omega 'r}{2}\right\}
\nonumber\\ \nonumber\\
&&+\sigma R_{gg}j^\theta_{(n)}\omega r^2
\sin\theta\cos\theta e^{-\Phi}+\sigma\beta T_{,\phi}\ .
\end{eqnarray}

Assume for simplicity of calculations that a superconducting ring
(thin disk of finite size with the inner radius $r_{in}$ and outer
one $r_{ex}$) is located in the equatorial plane $(\theta
=\pi/2)$, symmetrically around the axis of rotation. Then the
nonvanishing radial heat flow through superconducting ring may
create electric current of thermoelectric origin: $j_{(n)r}=
\sigma\beta e^{-\Phi}\left(e^\Phi T\right)_{,r}$. Then as a
consequence of azimuthal component of Ohm's law~(\ref{ohm_3}) the
azimuthal electric current will be created interior the
superconductor
\begin{eqnarray}
\label{j_3} j_{(s)\alpha} = -
\frac{2n_se}{m_s}\left[\hbar\partial_\alpha\vartheta
-\frac{2e}{c}A_\alpha\right]
\nonumber \\
= -R_{gg}\sigma^2\beta e^{-2(\Phi+\Lambda)}\left(e^\Phi
T\right)_{,r} \left\{\omega r -\omega r^2\Phi_{,r}+\frac{\omega
'r^2}{2}\right\}\ .
\end{eqnarray}

By integrating the equation~(\ref{j_3}) along the contour $r=R$,
which is in the bulk of the superconducting ring one can get
\begin{equation}
\label{mag_flux_final}
\Phi_b=n\Phi_0-\frac{\pi m_sc}{2n_se^2} R_{gg}\sigma^2\beta
e^{-2(\Phi+\Lambda)}\left(e^\Phi T\right)_{,r} \left\{\omega r
-\omega r^2\Phi_{,r}+\frac{\omega 'r^2}{2}\right\} _{|r=R}\ .
\end{equation}
Thus we considered the mechanism of production of stellar magnetic
field in superconductor with the heat flow.
The arguments on generation of magnetic field by the thermoelectric
effects in the conducting crust of the neutron star with the immense
gradient of temperature were extensively studied in papers
~\cite{bah83,uly86,wg96}. Since there is
uncertainty in microphysics of neutron stars we decided do not
provide here any evaluation for magnetic flux~(\ref{mag_flux_final})
for this toy model. The realistic model with the numerical evaluation
will be developed in our future work.

\section{Conclusion}
\label{conclusion}

Here we investigated the effect of the gravitation on the
electromagnetic properties of II-type superconductors and the
effects of uniform rotation on superconductors with a heat flow.
In particular we have shown that gravity can affect the
alternating voltage in II-type superconductor. The proposal is
presented which uses gravity to create an additional emf and
consequently may  change the superconducting Josephson effect
frequency. The frequency is becoming a function of the
gravitational field. Thus the results on the electromagnetic
properties of II-type superconductor embedded in gravitational
field could lead to a way towards measuring of the tiny general
relativistic effects in the earth conditions.

We have also derived the expression for magnetic flux through
rotating hollow type-I superconductor of cylindrical shape when a
temperature gradient is applied in radial direction between the
inner and outer boundaries of the sample. In principle, the
magnetic field due to the rotational effects on the electric
current in the superconducting state should be observable in this
type of laboratory experiment, although its magnitude is less
than flux quantum for many orders.

The mechanism of the generation of magnetic field inside rotating
neutron star due to the frame dragging effect has been considered.
A possibility of magnetic field production has been shown in the
toy model, where for the simplicity  superconducting matter has
been considered as the disc in the equatorial plane.

\section*{Acknowledgments}

BA and VK greatly thank TWAS and ICTP for financial support towards their
visit to India
 and acknowledge the warm hospitality at the IUCAA where
the work has been done. BA acknowledges the partial financial
support from NATO through the reintegration grant EAP.RIG.981259.
Financial support for this research is partly provided by the
UzFFR (project 1-06) and projects F2.1.09 and F2.2.06 of the
UzCST.

\end{document}